\documentclass[a4paper]{jpconf}

\usepackage{graphicx}
\usepackage{amsmath,amsfonts,amssymb}

\def\nn{\nonumber}
\def\mbb{\mathbb}

\newcommand{\mc}{\mathcal}

\newcommand{\bea}{\begin{eqnarray}}
\newcommand{\eea}{\end{eqnarray}}
\newcommand{\bei}{\begin{itemize}}
\newcommand{\eei}{\end{itemize}}
\newcommand{\vo}{{\cal V}}

\def\nn{\nonumber}

\def\be{\begin{equation}}
\def\ee{\end{equation}}
\def\ben{\begin{enumerate}}
\def\een{\end{enumerate}}

\begin{document}

\title{Global D-brane models with stabilised moduli and light axions}

\author{Michele Cicoli}

\address{ICTP, Strada Costiera 11, Trieste 34014, Italy and INFN Sezione di Trieste, Italy}

\ead{mcicoli@ictp.it}

\begin{abstract}
We review recent attempts to try to combine global issues of string compactifications,
like moduli stabilisation, with local issues, like semi-realistic D-brane constructions.
We list the main problems encountered, and outline a possible solution which
allows globally consistent embeddings of chiral models. We also argue that this stabilisation mechanism leads to an axiverse.
We finally illustrate our general claims in a concrete example where the Calabi-Yau manifold is explicitly described by toric geometry.
\end{abstract}

\section{Introduction}

Two longstanding problems of string compactifications are moduli stabilisation
and the derivation of realistic D-brane models.
Type II theories are promising to address these issues because of local sources.
On one side, D-branes provide non-Abelian gauge bosons whereas,
on the other side, local sources allow to evade no-go theorems which
prevent the turning on of background fluxes. In turn, these fluxes generate a potential for most of the moduli.
Moreover, in this \emph{brane-world scenario}, finding a solution to these two crucial problems
seems to be somewhat easier since moduli fixing is a \emph{global} issue whereas model building is a \emph{local} issue.
Thus one can hope that, at least at leading order, a separate study of the two issues makes sense.

So far, people have found viable mechanisms to fix the moduli especially in type IIB where
the backreaction of the fluxes is very mild and semi-realistic D-brane models can also be constructed.
It is therefore now time to combine the two solutions. The first attempts to do so
have shown that the two issues are not completely decoupled since one generically faces the following problems:
\bei
\item Tension between moduli fixing by non-perturbative effects and chirality \cite{blumenhagen};
\item Tension between moduli fixing by non-perturbative effects and the cancellation of Freed-Witten anomalies \cite{BBGWCKMW};
\item Possible shrinking of various divisors induced by D-terms (especially the four-cycles supporting the visible sector) \cite{BBGWCKMW,CKM}.
\eei
The main goal is then to overcome these difficulties keeping control over the 4D supergravity
theory and stabilising the moduli inside the K\"ahler cone.
Another non-trivial requirement is getting, at the same time, also an interesting
particle phenomenology and cosmology.

In this review, we focus on type IIB/F-theory compactifications with D3/D7-branes and O3/O7-planes
where the compact Calabi-Yau three-fold can be explicitly described in terms of toric geometry \cite{CMV}.
We perform a choice of brane set-up and world-volume fluxes that yields a chiral MSSM-like model.
We also check the global consistency of our model by
ensuring the cancellation of D7-tadpoles, torsion charges and Freed-Witten anomalies
(leaving enough space in the D3-tadpole to turn on three-form fluxes).
The K\"ahler moduli are fixed within the K\"ahler cone and the regime of validity of the
effective low-energy theory in a way compatible with chirality.
Moreover, the D-terms do not induce the shrinking of any divisor.\footnote{For the global embedding of models with
fractional branes at del Pezzo singularities see \cite{GlobalQuiver}.}
We also show how this moduli stabilisation mechanism (the LARGE Volume Scenario) leads
naturally to an axiverse whose main feature is the presence of many axions with exponentially small masses \cite{LVSaxiverse,Axiv}.
Our model is also able to reproduce a visible sector gauge coupling of the correct size and
interesting phenomenological scales like TeV-scale supersymmetry and an intermediate scale
decay constant for the QCD axion (realised as a local closed string mode, see \cite{Ringwald:PASCOS})
for natural values of the underlying parameters.

\section{K\"ahler moduli stabilisation}
\label{TmodStab}

The closed string moduli of type IIB compactifications are the axio-dilaton $S$,
the complex structure moduli $U_{\alpha}$, $\alpha=1,...,h^{2,1}$ and the K\"ahler moduli $T_i=\tau_i+{\rm i}\, c_i$,
where $\tau_i={\rm Vol}(D_i)$ and $c_i=\int_{D_i}C_4$, $i=1,...,h^{1,1}$.
The background fluxes $G_3 = F_3 + {\rm i}S H_3$ generate a tree-level superpotential
$W_{\rm tree}(S,U)$ which fixes $S$ and $U$ at $D_{S,U}W=0$. On the other hand, due to the \emph{no-scale structure},
the $T$-moduli are flat at tree-level, and so when one studies K\"ahler moduli stabilisation,
the $S$ and $U$-moduli can be fixed at their flux-stabilised values. Thus we can consider
$W_0 = \langle W_{\rm tree}\rangle$ and $K_{\rm tree} = - 2 \ln\vo$, where $\vo$ is a function of
the $T$-moduli which gives the volume of the internal manifold.
There are several sources for K\"ahler moduli stabilisation:
\be
V = V_D+V_F^{\rm tree} + V_F^{\rm pert} + V_F^{\rm np}\,.
\ee
In the previous expression $V_D \sim \mc{O}(1/\vo^2)$ is D-term potential (generated by fluxes on D7-branes),
$V_F^{\rm tree} \sim \mc{O}(1/\vo^2) = 0$ is the tree-level F-term potential which vanishes due to the no-scale structure,
$V_F^{\rm pert}\lesssim \mc{O}(1/\vo^3)$ contains perturbative effects which come from $\alpha'$ and $g_s$ corrections to $K$,
and finally $V_F^{\rm np}\lesssim \mc{O}(1/\vo^3)$ is derived from non-perturbative corrections to $W$
(E3-instantons or gaugino condensation on a D7-stack).
Our strategy will be to focus on the large volume limit, $\vo\gg 1$, and study the
behaviour of $V$ at each order in an inverse volume expansion.
At leading order in $1/\vo$, we look for supersymmetric solutions by imposing $V_D = 0$.
At subleading order, we instead minimise $V_F$ leading to stable vacua which break supersymmetry.

Let us now briefly discuss the main problems faced when trying to freeze the $T$-moduli
in a way consistent with the presence of chirality and world-volume fluxes.

\subsection{Non-perturbative effects and chirality}
\label{NPeffectsANDchirality}

There is a well known tension between K\"ahler moduli fixing by
non-perturbative effects and chirality \cite{blumenhagen}. This is because chirality is induced
by non-zero fluxes on intersections of branes, implying that the visible sector
has $\mc{F} \neq 0$. Moreover, if there are chiral modes on the intersection
between the cycle supporting non-perturbative effects $T_{\rm np}$ and the visible sector,
the prefactor $A$ of the non-perturbative superpotential
$W_{\rm np} = A \,e^{- a T_{\rm np}}$ depends on visible modes $\phi$.
In order to preserve the visible gauge group, $\langle\phi\rangle$ has to vanish,
implying $A = 0$ and no contribution to $W_{\rm np}$.

This tension constraints the flux choice: there has to be no chirality at the intersection
between the cycles supporting non-perturbative effects and the visible sector.
As pointed out in \cite{CKM}, this implies that the best place for non-perturbative effects
is a `diagonal' del Pezzo divisor since it is rigid and it has only self intersections.

\subsection{Non-perturbative effects and Freed-Witten anomaly}

As shown in \cite{FW}, one has to turn on a half-integer flux
on any \emph{non-spin} four-cycle $D$ ($c_1(D)$ is odd) to cancel Freed-Witten (FW) anomalies:
$F= f^i\eta_i + \frac 12 c_1(D)$, with $f^i\in\mbb{Z}$ and $\eta_i \in H^2(D,\mbb{Z})$.
Moreover, an $O(1)$ E3-instanton contributes to $W_{\rm np}$ if the E3
wraps a rigid cycle which is transversally invariant under the orientifold
and has total flux $\mc{F} = F - B = 0$ (similar considerations apply also for a gaugino condensation stack).
However, if the cycle is non-spin, the cancellation of FW anomalies requires $F \neq 0$.
One can then cancel $F$ by a proper choice of $B$ but once $B$ is fixed to cancel a half-integral flux on stack $a$,
generically $\mc{F} \neq 0$ on a second non-spin stack $b$ (unless they do not intersect) \cite{BBGWCKMW}.
Hence the cancellation of FW anomalies generically prevents to have more than
one non-perturbative effect to fix the $T$-moduli.

Our strategy to overcome this problem is the simplest one: fix the $T$-moduli by only one non-perturbative effect.
This leads to the LARGE Volume Scenario (LVS) where the presence of just a single diagonal del Pezzo with
non-perturbative effects is enough to ensure the stabilisation of the Calabi-Yau volume by setting $W_0\sim\mc{O}(1)$
and exploiting $\alpha'$-corrections to $K$ \cite{LVS}.

\subsection{D-term problem}
\label{DtermProbl}

As we have seen, world-volume fluxes are necessary for chirality and to
cancel worldsheet anomalies. In turn, these fluxes generate
Fayet-Iliopoulos (FI) terms of the form $\xi_a\propto\int_{D_a} J \wedge \mc{F}_a= k_{ajk} \mc{F}_a^k t^j$.
If the VEV of all charged fields vanishes, the D-term conditions imply $\xi_a = 0$,
generically forcing the shrinking of some four-cycles \cite{BBGWCKMW,CKM}. 
Let us explain this problem more in detail. 
In general, $\xi_a=0$ is a system of homogeneous linear equations in all the $h^{1,1}$ K\"ahler moduli.
However, if the non-perturbative effects are supported on $n_{\rm np}$ diagonal del Pezzo divisors,
these would not enter in $\xi_a = 0$ since they have no chiral intersections. Thus the generic
number of unknowns is $n = h^{1,1}-n_{\rm np}$. If the matrix of the system $\xi_a = 0$
has $d = n$, the only solution is the one where all the $t$'s collapse to zero size.
In order to avoid this shrinking, one has to choose the brane configuration and the fluxes such that
$d < n$ \cite{CMV}. In this case, the D-terms fix $d$ directions and leave $(n-d)$ flat directions.
If $n - d = 1$, all the $t$'s have to be of the same size which must be small
to obtain the right value of the visible gauge coupling: $g^{-2}\sim t^2$.
Thus in this case one cannot realise the LVS since this is characterised by an exponentially large volume.
The only way to get LVS is to ensure that $n - d = 2$.
If $d = 1$, the minimal $n$ to allow for LVS is then $n = 3$, leading to models with $h^{1,1} = 4$ for $n_{\rm np}=1$.

\section{The axiverse and moduli stabilisation}
\label{Axiverse}

The previous analysis has taught us that the K\"ahler moduli have to be fixed
by a combination of different effects (for $W_0\sim\mc{O}(1)$). Let us now study
what is the implication of this stabilisation scheme for the axions of the effective field theory:
\bei
\item $d$ combinations of $T$-moduli are fixed by the leading D-term potential,
and so $d$ axions get eaten up by anomalous $U(1)$s;
\item $n_{\rm np}$ `diagonal' del Pezzo divisor are fixed by non-perturbative effects,
and so the corresponding axions become heavy since they obtain the same mass of order $m_{3/2}$;
\item The remaining $n_{\rm ax} = h^{1,1}-n_{\rm np}-d\geq 2$ moduli have to be fixed perturbatively,
and so the corresponding axions remain massless. In particular,
the volume is frozen by $\alpha'$ corrections to $K$ whereas all the remaining moduli
are stabilised by subleading $g_s$ corrections to $K$.
\eei
This analysis implies that chiral models with the visible sector in the geometric regime
have always $n_{\rm ax} \geq 2$ light axions with possibly a very large $n_{\rm ax}$ for $h^{1,1}\sim \mc{O}(100)$.
One axion has to be the QCD axion and develops a mass via QCD instanton effects. On the other hand,
all the other axions get a tiny mass via higher order non-perturbative effects.
This is the typical picture of the \emph{axiverse} where there is a plethora of ultra-light axions
whose mass spectrum is logarithmically hierarchical \cite{LVSaxiverse,Axiv}.

\section{Explicit example}

Let us now illustrate our general claims in an explicit example taken from \cite{CMV} where the internal manifold
is built by means of toric geometry.

\subsection{A K3-fibred Calabi-Yau}

The Calabi-Yau three-fold is a hypersurface in a 4D toric ambient variety. Its weight matrix is:
\be
 \begin{array}{|c|c|c|c|c|c|c|c||c|}
 \hline  z_1 & z_2 & z_3 & z_4 & z_5 & z_6 & z_7 & z_8 & D_\textmd{CY} \tabularnewline \hline \hline
    1  &  1 &   1 &   0 &   0 &   0 &   1 &   4  & 8 \tabularnewline\hline
    1  &  1 &   0 &   0 &   0 &   1 &   0 &   3  & 6 \tabularnewline\hline
    0  &  1 &   1 &   1 &   0 &   0 &   0 &   3  & 6 \tabularnewline\hline
    0  &  1 &   0 &   0 &   1 &   0 &   0 &   2  & 4 \tabularnewline\hline
 \end{array}\,.
\label{weightmatrix}
\ee
The toric variety is the moduli space of the corresponding gauged linear sigma model
where the size of the FI terms $\xi_i$, $i=1,...,4$, is given by the K\"ahler moduli.
Thus this manifold has $h^{1,1} = 4$ (and $h^{1,2} = 106$). The Calabi-Yau is a divisor of the
ambient variety described by the last column in (\ref{weightmatrix}). The topological data are obtained from PALP \cite{PALP}:
\bei
\item Basis of $H_4({\rm CY},\mbb{Z})$: $\qquad\Gamma_1 = D_7\,,  \qquad \Gamma_2 = D_2 + D_7\,,  \qquad \Gamma_3 = D_1\,, \qquad \Gamma_4 = D_5$\,;

\item Intersection form: $\qquad I_3 = 2 \Gamma_1^3 +4\Gamma_2^3 +4\Gamma_4^3 +2 \Gamma_2^2\Gamma_3 -2 \Gamma_4^2\Gamma_3$\,.
\eei
The internal manifold is a K3 fibration with a diagonal del Pezzo \cite{CKM}. The K3 fibre is $D_1$
whereas the `diagonal' dP$_7$ corresponds to $\Gamma_1 = D_7$.
There are also three other rigid (but not del Pezzo) divisors: $D_4$, $D_5$ and $D_6$.
Expanding the K\"ahler form as $J =\sum_{i=1}^4 t_i \Gamma_i$, the Calabi-Yau volume becomes
$\vo = \frac13 \left[ 2 t_2^3 +3 t_2^2 t_3  + t_4^2( 2 t_4-3 t_3)+t_1^3\right]$ while the
volume of $D_7$ is simply $\tau_7 = t_1^2$. The K\"ahler cone is instead defined as:
\be
r_1\equiv- t_1>0\,, \qquad r_2\equiv t_1 + t_2 + t_4>0\,, \qquad r_3\equiv t_3 - t_4>0\,, \qquad r_4\equiv-t_4>0\,.
\ee
The reason why we focused on K3-fibred manifolds is because they are promising
for particle physics and cosmology. In fact, they can lead to anisotropic compactifications
that realise scenarios with supersymmetric large extra dimensions \cite{StringySLED},
light hidden $U(1)$s with kinetic mixing with the photon \cite{HiddenPhotons} and
natural string models of quintessence \cite{Quintessence}. Moreover, compactifications where the size
of the overall volume is controlled by more than one divisor are the starting point to build
both single-field \cite{FI} and multi-field \cite{NG} inflationary models.

\subsection{Model building with D7-branes}

Let us briefly review how semi-realistic models can be built via D7-branes wrapped around
smooth four-cycles. In order to get an $\mc{N}=1$ low-energy theory, one has to consider
orientifold projections $O = (-1)^{F_L}\Omega_p\sigma$ where $\sigma$ is a holomorphic involution.
By construction, we shall always consider divisors $D$ which are transversally invariant, and so
$N_a$ D7-branes (plus their images) wrapped around $D$ support an $Sp(2N_a)$ gauge group.
If we switch on a diagonal flux $\mc{F}$ on the branes wrapping $D$, the gauge group breaks down
to $Sp(2N_a) \to SU(N_a) \times U(1)$, so obtaining phenomenologically interesting unitary groups.
The diagonal $U(1)$ instead becomes massive via the St\"uckelberg mechanism.
Moreover, the D7-brane flux generates chiral modes. The number of chiral zero-modes
in symmetric and antisymmetric $U(N_a)$ representations is:
\be
I_a^{(S,A)}=\mp\frac12 \int_{{\rm CY}} [D7_a] \wedge [O7] \wedge {\cal F}_a  -  \int_{{\rm CY}} [D7_a] \wedge [D7_a] \wedge \mc{F}_a\,.
\ee
On the other hand, chiral zero-modes in bi-fundamental representations $(N_a, \overline{N}_b)$ and $(N_a,N_b)$
live at the intersection between two different stacks ($a$ and $b$). Their number is given by:
\be
 I_{a\bar{b}}= \int_{{\rm CY}} [D7_a] \wedge [D7_b] \wedge ({\cal F}_a - {\cal F}_b)\,, \qquad
 I_{ab}=\int_{{\rm CY}} [D7_a] \wedge [D7_b] \wedge ({\cal F}_a + {\cal F}_b)\,. \nn
\ee

\subsection{Charge cancellation}

In order to have a globally consistent compactification, the homological charges have to be cancelled.
These are encoded into the following quantities:
\be
\Gamma_{D7} = [D7] + [D7]\wedge {\cal F} + [D7]\wedge\left( \frac12 \,{\cal F} \wedge {\cal F}  + \frac{\chi(D7)}{24} \right),
\quad \Gamma_{O7} =  -8[O7]  + [O7]\wedge \frac{\chi(O7)}{6} \,. \nn
\ee
The D7-charge cancellation requires $\Sigma_{D7} [D7] = 8[O7]$, while the total D5-charge
is zero by construction since all branes and image-branes wrap the same divisor and $\mc{F}'=-\mc{F}$.
The total D3-charge gets contributions from the geometry and the fluxes. We shall not explicitly
turn on $H_3$ and $F_3$ but we shall check that the D3-tadpole leaves enough space to include them.
Also K-theoretic torsion charges must sum to zero. This is equivalent to the absence of $SU(2)$
gauge anomalies on any probe $Sp$-brane \cite{Uranga}.

\subsection{Orientifold projection and D7-brane stacks}

We perform the following choice for the holomorphic involution $\sigma$: $z_8 \mapsto -z_8$.
This gives an O7-plane at $z_8 = 0$ but no O3-planes. All the divisors are invariant, implying that
$h^{1,1}_- = 0$ and $h^{1,1}_+= h^{1,1}$.
In order to cancel the D7-charge of the O7, we need to have a D7-configuration on the divisor
class $8[D_8]$ which is described by the polynomial $\eta^2-z_8^2\chi=0$. This
corresponds to a \emph{Whitney brane} with double intersection with the O7.
If we want to have different stacks, this polynomial has to factorise.
Choosing $\eta$ and $\chi$ of the form $\eta=z_i^m\tilde{\eta}$ and $\chi=z_i^{2m}\tilde{\chi}$,
the initial polynomial reduces to $z_i^{2m}\left(\tilde{\eta}^2-z_8^2\tilde{\chi}\right)$.
This corresponds now to one $Sp(2m)$ stack along $z_i = 0$ plus a Whitney brane.
More generally, requiring $N_a$ branes on $D_4$, $N_b$ on $D_5$, $N_{k3}$ on $D_1$
and $N_{gc}$ on $D_7$ (plus their images), we get $z_1^{2N_{k3}} z_4^{2N_a} z_5^{2N_b} z_7^{2N_{gc}}\left(\tilde{\eta}^2-z_8^2\tilde{\chi}\right)$.
One can easily check that there is no further factorisation if $N_{gc}\leq 4$, $N_{gc}+N_{k3}\leq 4+N_a$
and $N_a -N_b\leq N_{gc}$. In order to have $\mc{F}_{gc} = 0$ on the gaugino condensation stack $N_{gc}$,
we need to choose $B = F_{gc}$, where $F_{gc}$ is the FW flux.
Finally the K-theory constraints are satisfied if $N_b$ is an even number.

\subsection{Example with one D-term}

Let us perform an explicit choice of brane set-up which satisfies all the previous constraints:
$N_a=3$, $N_{k3}=1$, $N_{gc}=3$ and $N_b=0$ with world-volume fluxes $\mc{F}_a^{\sigma}=\mc{F}_a$, $\sigma=1,2,3$, (diagonal flux) on $D_4$
and $\mc{F}_{k3}=0$ on $D_1$. The gauge group is broken to:
\be
 Sp(6) \times SU(2) \times Sp(6)\rightarrow SU(3)\times U(1) \times SU(2) \times Sp(6)\rightarrow  SU(3) \times SU(2) \times Sp(6)\,. \nn
\ee
Given that $N_b=0$ and $\mc{F}_{k3}=0$, only one D-term is generated.\footnote{For a GUT-like example
with gauge group $SU(5) \times U(1) \times Sp(8)$ and two D-terms see \cite{CMV}.}
The chiral modes of the model are $I^{(A)}_a = 2\beta_a-\nu$, $I^{(S)}_a = -2\beta_a+3\nu$,
$I_{ak3} = 2\alpha_a$, $I_{aW} = 4(4\beta_a-\alpha_a) + 8\,\nu$, $I_{k3W} = 0$ and $I_{agc}  = \nu$
where $\alpha_a$, $\beta_a$ and $\nu$ are integral combinations of flux numbers.
A possible choice of flux numbers consistent with all our requirements is $\alpha_a=1$, $\beta_a=-1$ and $\nu=0$.
Then, the total D3-charge turns out to be $Q_{(D3)}^{\rm tot} = -606$ and
the only non-zero chiral intersections are $I^{(A)}_a = -2$, $I^{(S)}_a = 2$, $I_{ak3} = 2$ and $I_{aW} =  -20$.

\subsection{D-term potential}

The model involves just one non-trivial FI-term: $\xi_a \propto \int_{{\rm CY}} [D7_a] \wedge J\wedge {\cal F}_a
 = \left[ (\beta_a-\alpha_a)( r_1+ r_2) + 2\alpha_a r_3\right]$.
Substituting our flux choice, the solution to $\xi_a=0$ is $\tau_4=3\left(\tau_1-\tau_5\right)-\tau_7$.
Plugging this relation into the Calabi-Yau volume, this reduces to
$\vo=\frac 16 \sqrt{\tau_1 - \tau_5} \left(10 \tau_1 - \tau_5\right) - \frac 13 \,\tau_7^{3/2}$.
Since $\alpha_{\rm vis}^{-1}=\tau_4$ and the del Pezzo divisor $\tau_7$ will be fixed at small size,
also the combination $\tau_s\equiv\left(\tau_1-\tau_5\right)$ has to be fixed small. Defining the
`big' cycle as $\tau_b\equiv\left(10\tau_1-\tau_5\right)/2$, the volume form further simplifies to
$\vo=\frac 13 \left( \sqrt{\tau_s}\, \tau_b - \,\tau_7^{3/2}\right)$.

\subsection{F-term potential}

The F-term potential depends on $\tau_s$, $\tau_b$ and $\tau_7$.
The leading order contributions to the potential come from $\alpha'$ and non-perturbative corrections
which give rise to (at leading order in $1/\vo$):
\be
V\simeq 2\pi ^2 A^2 \frac{\sqrt{\tau_7}}{\vo}\, e^{-\pi  \tau_7}
- 2\,\pi A  W_0 \,\frac{\tau_7}{\vo^2}\,e^{-\frac{\pi  \tau_7}{2}}+\frac{3 W_0^2 \hat\xi }{4 \vo^3}\,.
\label{V}
\ee
This potential depends only on $\vo$ and $\tau_7$, and so there is one flat direction left over.
Equation (\ref{V}) is the typical LVS potential which has a SUSY-breaking minimum at
$\vo\simeq\left[ W_0 \,\sqrt{\tau_7}/\left(2 \pi  A\right)\right]\,e^{\frac{\pi  \tau_7}{2}}$ and
$\tau_7 \simeq \left(3 \xi/2\right)^{2/3} g_s^{-1}$.
For natural values of the parameters, $W_0= 1$, $A=0.1$ and $g_s=0.05$,
one finds $\langle\tau_7\rangle\simeq 16$ and $\langle\vo\rangle \simeq 10^{12}$,
justifying the validity of our approximations. This choice gives TeV-scale supersymmetry since the
gravitino mass is $m_{3/2}=  \sqrt{g_s/(8\pi)} W_0 M_P/\vo \simeq 100$ TeV
while the soft terms generated by gravity mediation are of the order $M_{\rm soft}\simeq m_{3/2}/\ln\left(M_P/m_{3/2}\right)\simeq 3$ TeV.
The string scale is intermediate, $M_s \simeq M_P/\sqrt{4 \pi \vo}\simeq 10^{11}$ GeV,
giving a perfect decay constant for local axions: $f_{a}\sim M_s$ \cite{LVSaxiverse,Ringwald:PASCOS}.
String loop corrections can lift the remaining flat direction at subleading order, stabilising $\tau_s$ small
($\tau_s\simeq 30$) and $\tau_b$ large ($\tau_b\simeq 10^{11}$) well inside the K\"ahler cone \cite{CMV}.
Thus the Calabi-Yau is very anisotropic.
Note that a slightly different value of $g_s$ ($g_s = 0.02$) gives
a huge value of the volume, $\vo\simeq 10^{29}$, which leads to
stringy scenarios with $M_s \simeq 2$ TeV
and two micron-sized extra dimensions \cite{StringySLED}.

\section{Conclusions}

We outlined a general strategy to combine K\"ahler moduli stabilisation
with chiral D-brane models in type IIB compactifications.
We presented a concrete chiral model by means of toric geometry. This allowed to make a specific choice of brane set-up and fluxes
that gave rise to an MSSM-like model where several consistency constraints have been checked.
We also computed the scalar potential and minimised it,
obtaining interesting phenomenological scales. This is the first
explicit realisation of an LVS model.
Finally we argued that this moduli fixing mechanism leads to an axiverse.
Crucial issues that we leave for future research are: $(i)$ an analysis of background fluxes to fix the $S$ and $U$-moduli;
$(ii)$ the derivation of the right spectrum and Yukawa couplings;
$(iii)$ a systematic search through the PALP list of Calabi-Yau three-folds.

\ack

I would like to thank Mark Goodsell, Christoph Mayrhofer, Andreas Ringwald and Roberto Valandro
for the fruitful collaboration on the topics covered in this review.

\end{document}